\documentclass[sigconf]{acmart}
\def \toolname {\emph{Compressor} }
\def \baseline {BiLSTM$_{\text{\emph{soft}}}$ }
\def \baselines {BiLSTM$_{\text{\emph{soft}}}$}
\usepackage{comment}
\usepackage{listings}
\usepackage{xcolor}
\usepackage{balance}
\usepackage{tcolorbox}
\usepackage{listings}
\definecolor{mGreen}{rgb}{0,0.6,0}
\definecolor{mGray}{rgb}{0.5,0.5,0.5}
\definecolor{mPurple}{rgb}{0.58,0,0.82}
\definecolor{backgroundColour}{rgb}{0.95,0.95,0.92}

\lstdefinestyle{CStyle}{
    backgroundcolor=\color{backgroundColour},   
    commentstyle=\color{mGreen},
    keywordstyle=\color{magenta},
    numberstyle=\tiny\color{mGray},
    stringstyle=\color{mPurple},
    basicstyle=\footnotesize,
    breakatwhitespace=false,         
    breaklines=true,                 
    captionpos=b,                    
    keepspaces=true,                 
    numbers=left,                    
    numbersep=5pt,                  
    showspaces=false,                
    showstringspaces=false,
    showtabs=false,                  
    tabsize=2,
    language=C
}
\usepackage{enumitem}
\usepackage{multirow}
\usepackage[ruled,linesnumbered]{algorithm2e}
\usepackage{array}
\usepackage{arydshln}
\usepackage{bm}
\usepackage{url}
\usepackage[colorinlistoftodos]{todonotes}
\usepackage{subfigure}
\usepackage{makecell}
\SetAlFnt{\small}
\SetAlCapFnt{\small}
\SetAlCapNameFnt{\small}
\usepackage{algorithmic}

\algsetup{linenosize=\small}
\SetKwInput{KwInput}{Input}                
\SetKwInput{KwOutput}{Output}              

\newboolean{showcomments}
\setboolean{showcomments}{true}
\ifthenelse{\boolean{showcomments}}
{ }

\ifthenelse{\boolean{showcomments}}
{ }

\ifthenelse{\boolean{showcomments}}
{ }

\AtBeginDocument{%
  \providecommand\BibTeX{{%
    \normalfont B\kern-0.5em{\scshape i\kern-0.25em b}\kern-0.8em\TeX}}}

\copyrightyear{2022} 
\acmYear{2022} 
\setcopyright{acmlicensed}\acmConference[ASE '22]{37th IEEE/ACM International Conference on Automated Software Engineering}{October 10--14, 2022}{Rochester, MI, USA}
\acmBooktitle{37th IEEE/ACM International Conference on Automated Software Engineering (ASE '22), October 10--14, 2022, Rochester, MI, USA}
\acmPrice{15.00}
\acmDOI{10.1145/3551349.3556964}
\acmISBN{978-1-4503-9475-8/22/10}

\begin{document}

\title{Compressing Pre-trained Models of Code into 3 MB}

\author{Jieke Shi, Zhou Yang, Bowen Xu\textsuperscript{$\ast$}, Hong Jin Kang and David Lo}
\thanks{$\ast$ Corresponding author}

\affiliation{%
  \institution{School of Computing and Information Systems}
  \country{Singapore Management University}
}
\email{{jiekeshi, zyang, bowenxu.2017, hjkang.2018, davidlo}@smu.edu.sg}


\renewcommand{\shortauthors}{Shi et al.}


\begin{abstract}
  Although large pre-trained models of code have delivered significant advancements in various code processing tasks, there is an impediment to the wide and fluent adoption of these powerful models in software developers' daily workflow: these large models consume hundreds of megabytes of memory and run slowly on personal devices, which causes problems in model deployment and greatly degrades the user experience. 
  
  It motivates us to propose \emph{Compressor}, a novel approach that can compress the pre-trained models of code into extremely small models with negligible performance sacrifice. 
  Our proposed method formulates the design of tiny models as simplifying the pre-trained model architecture: searching for a significantly smaller model that follows an architectural design similar to the original pre-trained model. \emph{Compressor} proposes a genetic algorithm (GA)-based strategy to guide the simplification process. 
  Prior studies found that a model with higher computational cost tends to be more powerful. Inspired by this insight, the GA algorithm is designed to maximize a model's Giga floating-point operations (GFLOPs), an indicator of the model computational cost, to satisfy the constraint of the target model size. Then, we use the knowledge distillation technique to train the small model: unlabelled data is fed into the large model and the outputs are used as labels to train the small model.
  We evaluate \toolname with two state-of-the-art pre-trained models, i.e., CodeBERT and GraphCodeBERT, on two important tasks, i.e., vulnerability prediction and clone detection.
  We use our method to compress pre-trained models to a size ($3$ MB), which is $160\times$ smaller than the original size. The results show that compressed CodeBERT and GraphCodeBERT are $4.31\times$ and $4.15\times$ faster than the original model at inference, respectively.
  More importantly, they maintain $96.15\%$ and $97.74\%$ of the original performance on the vulnerability prediction task. They even maintain higher ratios ($99.20\%$ and $97.52\%$) of the original performance on the clone detection task. 
\end{abstract}




\maketitle

\section{Introduction}
\label{sec:intro}

Pre-trained models of code have achieved great success in various program understanding and generation tasks~\cite{feng2020codebert, guo2021graphcodebert, lu2021codexglue, wang2021codet5, ahmad2021unified}. 
However, these models are large-sized with billions of parameters and have low inference speed; thus they do not suit consumer-grade devices or applications that require strictly low latency. To put things in perspective, CodeBERT~\cite{feng2020codebert}, a state-of-the-art pre-trained model for processing multiple programming languages, has a total of 125 million parameters, which results in a large model size of 476 MB. 
When such a large model performs inference on a consumer-grade laptop, the response latency is as high as 1.5 seconds.\footnote{The results are collected by running a handful of examples with a token length of 400 on the vulnerability prediction task. The number of used CPU cores is limited to 8.} 
 According to the experience of the Visual Studio teams who have deployed integrated development environments (IDEs) to numerous users, Svyatkovskiy et al.~\cite{fastcompletion} suggest that a reasonable upper bound size for an IDE component or an editor plug-in is 50 MB, while a 3 MB model is preferred as it can be deployed even in severely restricted environments (e.g., low-end hardware for teaching students to code). Meanwhile, Aye et al.~\cite{aye2020sequence} argue that the inference latency should be reduced to 0.1 seconds as much as possible. Recently, researchers~\cite{9678546,fse_lightning} also try to minimize the response time of models for code.
 The pre-trained models of code are designed to serve coding tasks, promising to be a component in modern IDEs to assist developers in software development and maintenance.
However, the huge storage and runtime memory consumption, coupled with high inference latency, make these large models prohibitive for integration in modern IDEs.  
Therefore, it is imperative to reduce the size of these pre-trained models before deploying them.


To date, some approaches have been proposed to compress pre-trained models in natural language processing or other tasks~\cite{jiao2020tinybert, xu-etal-2020-bert, sun-etal-2020-mobilebert, zhang-etal-2020-ternarybert}. 
These existing studies fall under three types: model pruning, model quantization, and knowledge distillation. 
Model pruning works by replacing partial model parameters with zero~\cite{gordon-etal-2020-compressing} or removing architectural components (e.g., network layers or attention heads)~\cite{NEURIPS2019_2c601ad9}. 
Model quantization techniques convert the model's parameters from 32-bit floating-point numbers into low-bit fixed-point numbers~\cite{zadeh2020gobo}. 
Unfortunately, model pruning cannot compress CodeBERT and GraphCodeBERT to be less than 50 MB -- even if all network layers in the pre-trained model are removed, the model still leaves an embedding table of about 150 MB. 
Although model quantization can significantly reduce the model size by storing parameters using low-bit numbers, a compressed model is not faster in inference speed or consumes less CPU memory~\cite{10.1162/tacl_a_00413, zadeh2020gobo,pmlr-v139-kim21d}. 
Instead, specialized hardware or specialized low-level processing libraries are required at runtime.\footnote{Noted that quantization of 8-bit and above can reduce inference latency and memory consumption on CPUs with the support of PyTorch or TensorFlow libraries, but the size of the model after 8-bit quantization is bigger than 200 MB. Here we only discuss quantization below 8-bit.} We cannot expect IDE users' devices to have specialized hardware in addition to a consumer-grade CPU. A compressed model is expected to bring significant efficiency gains on CPUs in particular.

Different from model pruning and model quantization techniques, knowledge distillation promises to
produce a compressed model that is small enough and has efficiency gains on CPUs.
To transfer knowledge from a large model to a smaller model,
knowledge distillation trains a smaller \emph{student} model to mimic the behavior of the larger \emph{teacher} model~\cite{44873}. 
However, preserving knowledge after knowledge distillation is challenging. 
As pointed out by Cho et al.~\cite{Cho_2019_ICCV}, the \emph{gap in model capacity} between the large pre-trained models makes knowledge distillation unable to achieve good results. Vapnik et al.~\cite{vapnik1999nature} proposed model capacity to describe the complexity and the expressive power of a machine learning model or a set of functions. A tiny student model with low model capacity cannot absorb knowledge from large models effectively, which makes the tiny model greatly under-perform the large one~\cite{Mirzadeh_2020}. 
Compared with pre-trained models, tiny models have thinner and shallower network architectures, and hence, the model capacity is reduced accordingly. Therefore, carefully choosing an appropriate architecture with enough capacity for the small student model is fairly important.
Previous studies mostly focus on transferring the pre-trained model's knowledge to a tiny model with a pre-defined architecture. 
Instead, exploration of how to particularly design the architectures of tiny student models is limited in the literature.


It is challenging to obtain an appropriate model architecture that has enough capacity to preserve knowledge distilled from the large model. First, \emph{finding the appropriate architecture is essentially a combinatorial problem with massive search space.} There exist a large number of network structures to select from (e.g., CNN, LSTM), and these structures have their own architecture-related hyper-parameters that can be adjusted (e.g., number of network layers).
Second, \emph{it is computationally infeasible to evaluate each plausible candidate model by training and testing it, making it difficult to guide the search.}
As a result, an easy-to-compute and effective predictive metric is desired to be tailored for this difficult search problem.

Pursuant to the above challenges, we propose a novel technique, \emph{Compressor}, that efficiently finds an appropriate small student model that can effectively absorb the knowledge of the large model.
\emph{Compressor} first utilizes the architectural designs of pre-trained models to narrow the search space. In particular, it tries to find a student model that shares a similar architecture to the teacher model but has a smaller size, which we call the \emph{model simplification} problem.
Gao et al.~\cite{Residual} measure model capacity with the \emph{computational cost}, which is defined as the unit of Giga floating-point operations (GFLOPs). We adopt this measurement, i.e., the computational cost of a candidate model, as a predictive metric to guide the search.
We design a customized genetic algorithm (GA)-based algorithm and use the computational cost of a found solution as the fitness function. 
After searching with the GA-based algorithm, \emph{Compressor} applies knowledge distillation to teach the found small model. It feeds a number of unlabeled inputs into the pre-trained models and obtains the predictions for each input. Then, the knowledge (i.e., pairs of input and prediction) is used to train the small model.

To evaluate the effectiveness of \emph{Compressor}, we conduct experiments with two state-of-the-art pre-trained models of code, i.e., CodeBERT~\cite{feng2020codebert} and GraphCodeBERT~\cite{guo2021graphcodebert}. We experiment \toolname with a 3 MB model size and analyze the performance of compressed models on two downstream tasks that are important to assist daily software development activities: vulnerability prediction and clone detection. 
In particular, we use the \toolname to compress models to a size (3 MB) that is only 0.6\% of the original model size, the degradation in accuracy is negligible on both two tasks. On average, the models compressed by \toolname can maintain 96.95\% and 98.36\% of the accuracy of the original large models on both vulnerability prediction and clone detection tasks. Moreover, we compare the 3 MB models obtained by \toolname with the 7.5 MB models from Tang et al.~\cite{distillLSTM}, which is called \baselines. Results show that \toolname outperforms \baseline by 49.27\% and 87.29\% to reduce the accuracy loss of compressed models on both two tasks, respectively. Furthermore, compared with the large pre-trained models, the compressed models reduce the inference latency by 70.75\% and 79.21\% on vulnerability prediction and clone detection tasks, respectively.
We also analyzed the efficiency of \emph{Compressor}. The average time spent on finding a tiny model with an appropriate architecture by \toolname is only $1.22$ seconds. Plus time consumption on training, \toolname brings a low additional time cost in compressing pre-trained models, which only takes 30.39\% and 37.06\% of the fine-tuning time on average to compress CodeBERT and GraphCodeBERT on two tasks. 
To conclude, this paper makes the following main contributions:
\begin{itemize}[leftmargin=*]
    \item Our work is the first to highlight the necessity of compressing pre-trained models of code and refine this problem into a model simplification process for pre-trained models.
    \item We propose \emph{Compressor}, a novel compression method via genetic algorithm (GA)-guided model simplification and knowledge distillation.
    \item We implement \toolname and evaluate it with CodeBERT and GraphCodeBERT across two downstream tasks. The results validate the feasibility of compressing pre-trained models with a negligible performance penalty.
\end{itemize}

The rest of this paper is structured as follows: Section~\ref{sec:preliminaries} covers the preliminary information of our work. Section~\ref{sec:method} presents the proposed \emph{Compressor}. Section~\ref{sec:exp_setup} describes the experimental setup. Section~\ref{sec:evaluation} analyses the experimental results. Section~\ref{sec:discussion} discusses the impacts of compressed model sizes and threats to validity. Section~\ref{sec:rel_work} presents related work, and Section~\ref{sec:conclusion} concludes our work.

\vspace{0.2cm}
\noindent \textbf{Replication Package}: The code and documentation, along with the obtained models, are made open-source for validating  reproducibility.\footnote{\url{https://github.com/soarsmu/Compressor.git}}

\section{Preliminaries}
\label{sec:preliminaries}

This section briefly introduces some preliminary information about this study, including pre-trained models of code, knowledge distillation, and the baseline we compare in the experiments.

\subsection{Pre-trained Models of Code}
\label{subsec:pre-trained}

Some pre-trained models for natural languages like BERT~\cite{DBLP:conf/naacl/DevlinCLT19} have recently demonstrated excellent performance in various language processing tasks. Inspired by the success of these language models, researchers have created pre-trained models of code~\cite{feng2020codebert, guo2021graphcodebert} that can benefit a broad range of programming language processing tasks. 

CodeBERT, which is proposed by Feng et al.~\cite{feng2020codebert}, can process multiple programming languages. CodeBERT shares the same model architecture as RoBERTa~\cite{liu2019roberta} and is pre-trained on the code search dataset provided by Husain et al.~\cite{CodeSearchNet}. This dataset consists of 6 million code functions and 2 million function-documentation pairs collected from open-source projects. CodeBERT has two pre-training tasks. One is called masked language modeling (MLM), which aims to predict the original values of some tokens that are masked out in an input. The other task is named replaced token detection (RTD); the model needs to determine whether a token in a given input is replaced. Experiment results have shown that in downstream tasks like the code-to-text generation or code search, which require the model to comprehend the code semantics, CodeBERT can yield superior performance.
GraphCodeBERT~\cite{guo2021graphcodebert} uses the same architecture as CodeBERT. Its difference with CodeBERT is that GrapCodeBERT additionally considers the inherent structure of code, i.e., data flow graph (DFG). GraphCodeBERT keeps the MLM training objective and discards the RTD objective. Instead, it designs two DFG-related tasks for learning representation from data flow: data flow edge prediction and node alignment. The former is to mask some edges in the data flow graphs and then let GraphCodeBERT predict those edges. Another task is to align the representation between source code and data flow, which lets GraphCodeBERT predict from which position in the source code a variable in the DFG is extracted. Results show that GraphCodeBERT outperforms CodeBERT on four downstream tasks~\cite{guo2021graphcodebert}.

There are some other pre-trained models of code. Two language-specific pre-trained models are  CuBERT~\cite{pmlr-v119-kanade20a} and C-BERT~\cite{CBERT}, which are trained on Python and C source code, respectively. CodeT5~\cite{wang2021codet5} is a Transformer-based model that considers identifier information during pre-training and can be used to address both code understanding and generation tasks. 
This paper focuses on compressing CodeBERT and GraphCodeBERT, considering their popularity and good performance. A recent study~\cite{lu2021codexglue} has empirically shown that CodeBERT and GraphCodeBERT demonstrate state-of-the-art performance across multiple code processing tasks.

\subsection{Knowledge Distillation}
\label{subsec:kd}
With the continuously increasing data and model size, Deep Neural Network (DNN) models have been successful on many tasks. 
However, deploying these advanced models is challenging, considering their large size and with limited power of deployed devices. 
Various model compression techniques have been proposed, and knowledge distillation is a representative of them~\cite{gou2021knowledge}.
Knowledge distillation compresses a large model by training a small model to mimic the behaviors of the model (i.e., produces the same output given the same input)~\cite{44873,NIPS2014_ea8fcd92,gou2021knowledge}.
The large model serves as a \emph{teacher}, and the small model learns to mimic the teacher as a \emph{student}. Many techniques have been proposed to show the effectiveness of knowledge distillation on different tasks, including image classification~\cite{Cho_2019_ICCV, liu2018multi}, sentiment analysis~\cite{distillLSTM}, speech recognition~\cite{yoon2021tutornet}, etc. 

According to how the compressed models are used, knowledge distillation techniques can be categorized into \emph{task-specific} distillation and \emph{task-agnostic} distillation. 
The former is to compress a pre-trained model that is fine-tuned on a specific task so that the compressed model can be applied to the downstream task directly. The latter is to compress the original pre-trained model directly and then fine-tune the compressed model on the downstream task~\cite{jiao2020tinybert,distillbert}. 
Task-agnostic distillation involves an extremely time-consuming and computational resource-hungry pre-training process. For instance, DistillBERT~\cite{distillbert} is trained on 8 16GB V100 GPUs for approximately 90 hours, while NASBERT~\cite{NASBERT} is trained on 32 NVIDIA P40 GPUs for 8 days. Such huge training costs can be prohibitive in model developments. Our study focuses on task-specific distillation as it incurs less time cost and has much lower computational resource requirements.

\section{Methodology}
\label{sec:method}

We explain the design of our proposed tool \emph{Compressor} in this section, including the formalization of the search problem, the GA-guided model simplification algorithm, and how to perform knowledge distillation to train the small models.

\subsection{Overview}
\label{subsec:overview}

\begin{figure}[!t]
  \centering
  \includegraphics[width=1.0\linewidth]{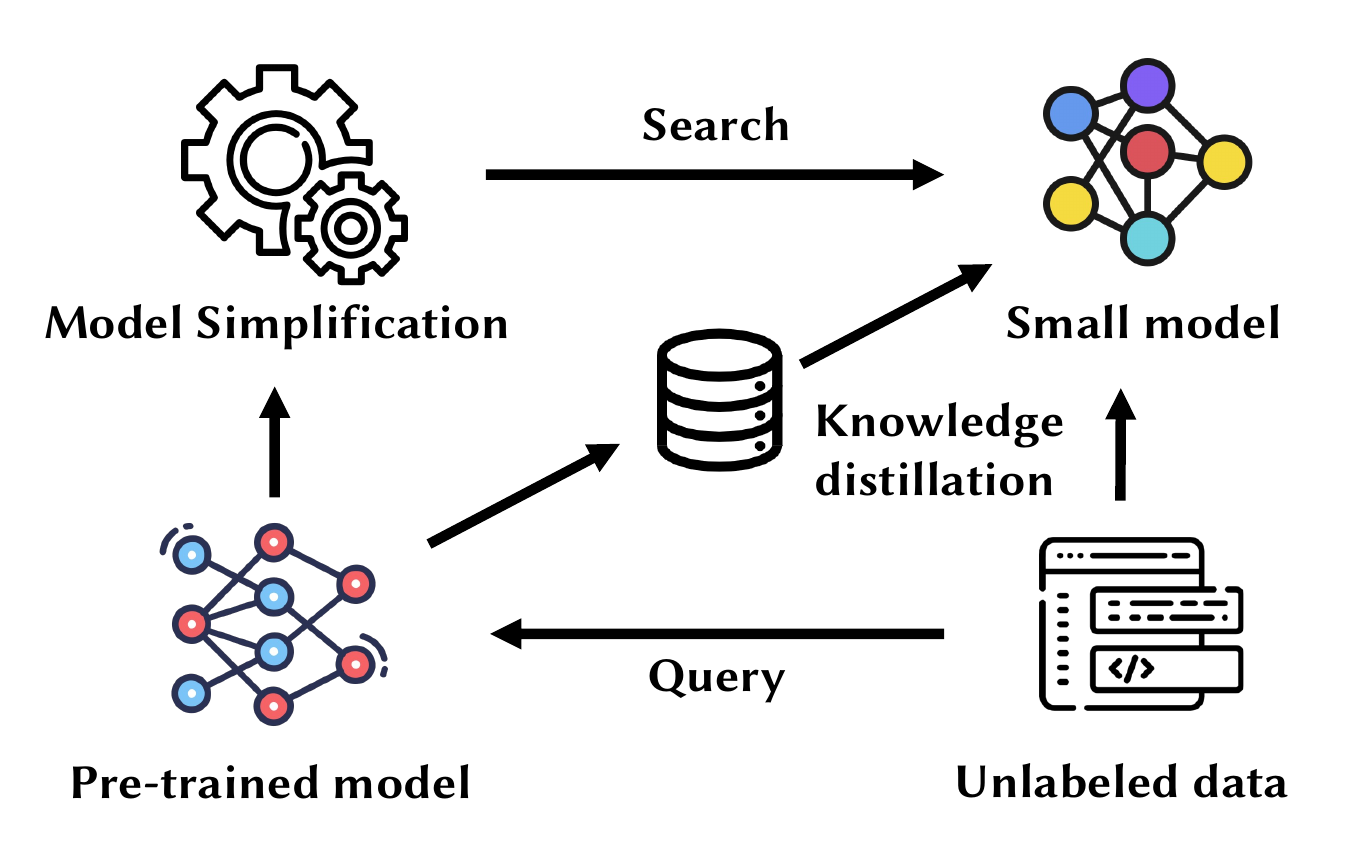}
  \caption{This figure illustrates an overview of the proposed model compression approach. The model simplification module analyzes the pre-trained models and searches for a small model using a GA-based strategy. Then we use unlabeled data to query the pre-trained model. Then the pre-trained model's outputs are used to train the small model, which is called knowledge distillation.}
  \label{online&offline}
\end{figure}

Knowledge distillation is a technique to compress a large model into a smaller one. First, we query the large model and obtain its outputs, which are treated as the `knowledge' of the large model. Then the distilled knowledge is used to train a smaller model so that the latter can maintain (most of) the behaviors of the large model but has a smaller size. However, the architecture of a small model can significantly impact the model performance after knowledge distillation~\cite{Cho_2019_ICCV}. The process of finding an optimal architecture for a small model is essentially a search problem: given a constraint on the model size, we want to search for a model that has the optimal performance after knowledge distillation.

To tackle the search problem, two main challenges need to be addressed. The first challenge is the huge search space with numerous plausible combinations. A model stacks many layers, each of which contains different numbers of parameters. Small changes to any element of the architecture may result in a new neural network that could produce largely different performance even when trained on the same dataset. Model developers usually put laborious engineering effort into finding an appropriate architecture for the tiny model, which is time-consuming and computing resource-hungry.
The second challenge is that the objective of this search problem, i.e., the performance of the tiny model after distillation, is very expensive to compute. It is impractical and infeasible to train and evaluate each model we find in the searching process. Therefore, an easy-to-compute and effective predictive metric is desired to be tailored for this difficult search problem.

In this work, we propose \emph{Compressor}, a novel approach that compresses pre-trained models of code via guided model simplification. \toolname is designed to solve the aforementioned two challenges.
To overcome the first challenge, \emph{Compressor} tries to follow the similar architectural design of the pre-trained models (i.e., the teacher models) to narrow the search space. In particular, it finds a student model that shares a similar architecture to the teacher model but has less width and depth, which we refer it as the \emph{model simplification} problem.
The candidate architectures are determined by a set of architecture-related hyperparameters, e.g., the number of layers, the network's dimensions, etc. 
To overcome the second challenge, we customize a genetic algorithm (GA)-based strategy to guide the process of model simplification, so as to find the promising architecture-related hyperparameters that can amplify the capacity of searched tiny models as much as possible and fit the given size limit meanwhile (presented in Section~\ref{sec:GA}). Finally, we adopt the knowledge distillation to obtain a well-performing small model (presented in Section~\ref{sec:distillation}).

\vspace{0.2cm}
\noindent\textbf{Application Scope.} 
A dozen of pre-trained models of code emerge across different academic venues. Most of them belong to BERT family as they uphold the same architecture as BERT~\cite{DBLP:conf/naacl/DevlinCLT19} or RoBERTa~\cite{liu2019roberta}, e.g., CodeBERT~\cite{feng2020codebert}, GraphCodeBERT~\cite{guo2021graphcodebert}, and CuBERT~\cite{pmlr-v119-kanade20a}. A few exceptions also maintain a Transformer-based architecture, e.g., CodeT5~\cite{wang2021codet5}, GPT-C~\cite{svyatkovskiy2020intellicode}, and CodeGPT~\cite{lu2021codexglue}. We argue that our study works out of the box for BERT-family pre-trained models, while requiring only minor modifications for other Transformer-based models. The current implementation of \toolname primarily focuses on the compression of CodeBERT and GraphCodeBERT due to their popularity and superior performance. However, our methodology can generalize to other Transformer-based models. We leave the evaluation of compressing other pre-trained models as future work (further discussion in Section~\ref{sec:extensions}).

\subsection{Problem Search Space}
\label{sec:design}

\begin{table}[t!]
  \centering
  \caption{Hyperparameters of BERT-family pre-trained models of code, and the search space of GA-guided model simplification including number of network layers ($L$), dimensionality of the network layers ($H$), number of attention heads ($A$), dimensionality of the feed-forward layers ($D$), and vocabulary size ($V$).}
  \renewcommand{\familydefault}{\sfdefault}\normalfont
  \begin{tabular}{@{}ccc@{}}
  \toprule
  Hyperparameter & \begin{tabular}[c]{@{}c@{}}Pre-trained\\ Models\end{tabular} & Search Space \\ \midrule
  $L$ & 12 & {[}1, 12{]}, interval=1 \\
  $H$ & 768 & {[}16, 768{]}, interval=16 \\
  $A$ & 12 & 1, 2, 4, 8 \\
  $D$ & 3072 & {[}32, 3072{]}, interval=32 \\
  $V$ & 50265 & {[}1000, 50000{]}, interval=1000 \\ \bottomrule
  \end{tabular}
  \label{tab:space}
  \end{table}



As pointed out by Liu et al.~\cite{liu-etal-2020-fastbert}, a student model that shares the same architecture as BERT, i.e., a bidirectional encoder architecture from transformers, can still achieve comparable or even outperform the large BERT model.
This indicates that sharing a similar architecture to the teacher model can upgrade the performance of the student model after knowledge distillation.
The size of a model in the BERT family relates to a set of hyperparameters.
We thoroughly analyze these hyperparameters and allow tuning of the following ones related to the model architecture to simplify a model: the number of network layers ($L$), the dimensionality of the network layers ($H$), the number of attention heads ($A$), the dimensionality of the feed-forward layers ($D$), and the vocabulary size ($V$). 
We limit the adjustable value of these hyperparameters to no larger than the corresponding ones in the pre-trained model, as shown in Table~\ref{tab:space}. Besides, in order to avoid the generation of invalid architectures with misaligned dimensions, hyperparameters are adjusted at specific intervals. The number of layers can be adjusted from 1 to 12. The range of the dimensionality of the network layers is from 16 to 768. The number of attention heads can be tuned as one size in $\{1, 2, 4, 8\}$. The dimensionality of the feed-forward layers can be tuned between 32 and 3072. The vocabulary size can be adjusted between 1000 and 50000.

\subsection{GA-guided Model Simplification}
\label{sec:GA}

Finding an appropriate simplified model under a model size constraint is essentially a combinatorial optimization problem, with an objective of finding the optimal combination of architecture-related hyperparameters that minimizes the difference between the small model size and a given size limit, and maximizes the capacity of the tiny model to achieve better performance in subsequent distillation.
However, it is non-trivial to identify a good combination of hyperparameters to adjust the structure of pre-trained models due to the large number of plausible hyperparameters. Although we have bounded such a search space, there are a total of 11,059,200 possible combinations of hyperparameters (possible candidates architectures of small models) to be tested. It is computationally infeasible to evaluate each plausible candidate model by training and testing it, making it difficult to guide the search.
Therefore, we propose a genetic algorithm (GA)-based strategy to guide the process of model architecture simplification.
Algorithm~\ref{algo:ga} shows the overview of how GA-guided model architecture simplification works. It first initializes the population (Line 1), and then performs genetic operators to generate new solutions (Line 2 to 11). Our GA-guided strategy computes the fitness function and keeps solutions with larger fitness values (Line 13). In the end, the algorithm returns the solution with the highest fitness value (Line 15 to 16).

\vspace{0.15cm}
\noindent\textbf{Chromosome Representation.}
In GA, a chromosome represents a solution to the target problem, consisting of a set of genes. In our study, each gene is a key-value pair of hyperparameter designation and specific value. Chromosomes store these key-value pairs in a dictionary data structure. For example, a chromosome can be \{$L$: 3, $H$: 512, $A$: 4, $D$: 1024, $V$: 10000\}, which represents a candidate architecture with 3 network layers of 512 dimensions and 4 attention heads, as well as 1024-dimensions feed-forward networks and 10000-sized vocabulary. Following common GA practice, we initialize a set of chromosomes by randomly setting the value of each key-value pair in the chromosome. These randomly initialized chromosomes, referred to as a population in GA terminology, provide a seed to launch iterations of evolution.

\vspace{0.15cm}
\noindent\textbf{Fitness Function.}
GA uses a fitness function to measure and compare the quality of chromosomes in a population. A higher fitness value indicates that the chromosome is closer to the target of this problem. 
As we discussed in Section~\ref{sec:intro}, a small model is expected to have enough model capacity to incorporate knowledge from the pre-trained model, thus guaranteeing comparable performance to large models at an extremely reduced model size. Disproportionate model capacity will make tiny models unable to well finish capturing and distilling knowledge.
Following Gao et al.~\cite{Residual}, we measure the model capacity by the computational cost in the unit of Giga floating-point operations (GFLOPs). GFLOPs count how many multiply and accumulate operations the model needs to make a forward pass. Larger GFLOPs mean that the model has a larger capacity.
Given a target model size, \toolname aims to find a small model that satisfies the model size and has the largest GFLOPs meanwhile. The fitness function is defined as follows:
\begin{equation}
\label{eq:fitness}
    Fitness(s) = GFLOP_s -|t_s - T|
\end{equation}
where $s$ is a chromosome, and $GFLOP_s$ is the computational cost of the architecture corresponding to $s$. $T$ is the given model size and $t_s$ is the size of the tiny model created according to $s$. $|t_s - T|$ is the difference between the model size of the current searched model and the target model size.

\vspace{0.15cm}
\noindent\textbf{Operators and Selection}
At each iteration, we employ two standard genetic operators, i.e., crossover and mutation, to produce new chromosomes. We apply crossover with a probability of $r$ (Line 7) and mutation with a probability of $1-r$ (Line 9). Given two chromosomes ($c_1$ and $c_2$), the crossover operator works as follows: we first randomly select a cut-off position $h$, and replace $c_1$'s genes after the position $h$ with $c_2$'s genes at the corresponding positions. The mutation operator also first randomly selects a cut-off position $h$, and then replace $c_1$'s genes after the position $h$ with a random value. 
After producing a new generation in one iteration, we merge them with the current population and perform a selection operator (Line 13). We always maintain a population of the same size and discard the chromosomes with lower fitness values.

\subsection{Distillation with Unlabeled Data}
\label{sec:distillation}
We advocate to use unlabeled data in knowledge distillation, as (1) Tang et al.~\cite{distillLSTM} and Jiao et al.~\cite{jiao2020tinybert} have shown that knowledge distillation without involving ground-truth labels can achieve better results; (2) labeled data is expensive and difficult to obtain, but unlabeled data can be easily scraped from the raw code of open-source projects in code hosting sites; and (3) a realistic scenario is that the deployed tiny model needs to be trained with newly generated data by users to harden performance, where no labeled data is available. 

Our study applies the knowledge distillation method introduced by Hinton et al.~\cite{44873}, where the student model learns from the teacher model's output probability directly. This method belongs to the task-specific distillation group that is mentioned in Section~\ref{subsec:kd}.
A recent study~\cite{decouple} confirms its effectiveness in transferring knowledge from a target large model to a small model across multiple tasks.
Specifically, we first feed unlabeled data into the larger pre-trained model and collect the output probability values. Then, we train the tiny model architecture identified by GA-guided model simplification following the objective introduced by Hinton et al.
We minimize the loss between the output of the pre-trained and tiny models by the following loss function:
\begin{equation}
  \mathcal{L} =-\frac{1}{n}\sum_{i}^{n} softmax(\frac{p_i}{T}) \log \left(softmax(\frac{q_i}{T}) \right) T^2
\end{equation}
In the above equation, $n$ is the number of examples in the training set. $p_i$ and $q_i$ are the outputs of the large model and the tiny model, respectively. $T$ is the temperature index to soften the softmax function introduced by Hinton et al.~\cite{44873}. Note that the pre-trained model producing $p_i$ is fixed during the distillation process while the small model producing $q_i$ is trained.

\begin{algorithm}[!t]
  \caption{GA-guided Model Simplification} 
  \label{algo:ga}
  \SetAlgoLined
  \KwInput{$c$: value ranges of architecture-related hyperparameters, \\
    $r$: crossover rate, $max\_iter$: maximum number of iterations, \\
    $child\_size$: number of generated children in each iteration}
  \KwOutput{$c'$: architecture-related hyperparameters}
  $population$ = random\_initialization($c$)\;
  \While{not exceed $max\_iter$}{
    {}
    $child\_list = []$ \;
    \While{$len(child\_lis) < child\_size$}{
        $p = random(0,1)$ \;
        \eIf{$p < r$}{
          $child$ = crossover($population$) \;
        }{
          $child$ = mutation($population$)\;
        }
        $child\_list$.append($child$)\;
    }
    $population$ = selection($population \cup child\_list$)\;
  }
  $c'$ = argmax(population); { \texttt{\# select the one with the highest fitness value}}\\
  \algorithmicreturn{ $c'$}
\end{algorithm}


\section{Experimental Settings}
\label{sec:exp_setup}

This section explains our experiment settings to evaluate the proposed \emph{Compressor}, including the implementation details, datasets, and evaluation criteria.

\subsection{Implementation and Configurations}
\label{sec:implementation}

\toolname is implemented in Python with around $2,000$ lines of code built upon HuggingFace Transformers. 
We borrow the work by Clark et al.~\cite{Clark2020ELECTRA} to calculate the FLOPs of a model and use APIs in PyTorch to calculate the model size.

In the experiments, the GA algorithm maintains a population of 50 candidate models. The crossover rate is set as 0.6.
The GA algorithm terminates when the number of iterations reaches 100. 
All the experiments are conducted on an Ubuntu 18.04 server with an Intel Xeon E5-2698 CPU, 504GB RAM, and 8 Tesla P100 GPUs. 

\subsection{Datasets and Tasks}
\label{sec:datasets}
To sufficiently evaluate \emph{Compressor}, we select two pre-trained models of code (CodeBERT~\cite{feng2020codebert} and GraphCodeBERT~\cite{guo2021graphcodebert}) that demonstrate state-of-the-art performance on a series of downstream tasks and use \emph{Compressor} to reduce their model sizes. We consider two important downstream tasks: vulnerability prediction and clone detection. 

\vspace{0.15cm}
\noindent\textbf{Vulnerability Prediction}.
This task aims to predict whether a given code snippet is vulnerable or not. Integrating the model for this task as an IDE component can greatly help developers identify potential code defects at an early stage. 
We use the dataset called Devign\footnote{\url{https://sites.google.com/view/devign}}, which was released by Zhou et al.~\cite{zhou_devign_2019}. 
Devign contains 27,318 functions extracted from two popular open-sourced C libraries called FFmpeg and Qemu. 
These functions are manually labeled as either containing vulnerabilities or not. 
Devign is included as a part of CodeXGLUE benchmark~\cite{lu2021codexglue} to investigate the effectiveness of CodeBERT for predicting vulnerabilities. 
Following the settings in CodeXGLUE, we use 2,732 examples for both validation and testing sets. 
We equally split the remaining examples into two mutually exclusive parts. One half is used to fine-tune the CodeBERT and GraphCodeBERT. The other half, whose labels are erased, is used to perform knowledge distillation and train the compressed models from scratch.

\vspace{0.15cm}
\noindent\textbf{Clone Detection}.
This task aims to decide whether two given functions are clones, i.e., equivalent in operational semantics. 
Clone detection can identify redundant implementations of the same functionalities. Finding code clones can help developers take actions to improve software quality, e.g., by code refactoring.
BigCloneBench~\cite{BigCodeBench} is a broadly-used benchmark for clone detection.\footnote{\url{https://github.com/clonebench/BigCloneBench}}
BigCloneBench is collected from various open-sourced Java projects.
It contains more than 6,000,000 pairs of cloned Java methods as well as 260,000 non-clone pairs. Following the settings of prior works~\cite{yang2022natural}, we randomly select $90,102$ examples for training and 4,000 for validation and testing to keep the experiment at a computationally friendly scale. Similar to what we have done to the vulnerability prediction dataset, the training data is also equally divided into labeled and unlabeled parts for obtaining the pre-trained and compressed models, respectively.

\vspace{0.15cm}
To fine-tune CodeBERT on the vulnerability prediction and clone detection tasks, we use the same hyperparameter settings adopted in the CodeXGLUE benchmark~\cite{lu2021codexglue}. To fine-tune GraphCodeBERT, we follow the same hyperparameter setting in the GraphCodeBERT paper~\cite{guo2021graphcodebert}. We follow the CodeXGLUE paper~\cite{lu2021codexglue} and Yang et al.~\cite{yang2022natural} to use accuracy as the evaluation metric to evaluate the model performance on downstream tasks.
Although the labeled training data we used is half of the full training set, all models achieve comparable results to the state-of-the-arts reported in the CodeXGLUE benchmark~\cite{lu2021codexglue} and the GraphCodeBERT paper~\cite{guo2021graphcodebert}. 
The statistics of datasets and the performance of fine-tuned models are presented in Table~\ref{tab:datasets}.

\begin{table}[t!]
    \centering
    \caption{Statistics of datasets and performance of the two pre-trained models.}
    \renewcommand{\familydefault}{\sfdefault}\normalfont
    \begin{tabular}{p{1.9cm}<{\centering}p{2.2cm}<{\centering}p{2cm}<{\centering}c}
    \toprule
    \multirow{2}{*}{Task} & \multirow{2}{*}{\begin{tabular}[c]{@{}p{2.2cm}<{\centering}@{}}Labeled/Unlabeled\\      Val/Test\end{tabular}} & \multirow{2}{*}{Model} & \multirow{2}{*}{Acc (\%)} \\
     &  &  &  \\ \midrule
    \multirow{2}{*}{\begin{tabular}[c]{@{}c@{}}Vulnerability\\  Prediction~\cite{zhou_devign_2019}\end{tabular}} & 10,927/10,927 & CodeBERT & 61.82 \\
     & 2,732/2,732 & GraphCodeBERT & 61.38 \\ \midrule
    \multirow{2}{*}{\begin{tabular}[c]{@{}c@{}}Clone\\  Detection~\cite{BigCodeBench}\end{tabular}} & 45,051/45,051 & CodeBERT & 96.20 \\
     & 4,000/4,000 & GraphCodeBERT & 96.62 \\ \bottomrule
    \end{tabular}
    \label{tab:datasets}
    \end{table}

\begin{table*}[t!]
    \centering
    \caption{Accuracy comparison of pre-trained models and compressed models. The size of each model is enclosed in parentheses. In the rows of Accuracy, the numbers in the parentheses correspond to the ratio of compressed model's accuracy to pre-trained models. In the rows of Drop, the numbers in the parentheses correspond to the ratio of relative improvements of \toolname compared to the baseline.}
    \renewcommand{\familydefault}{\sfdefault}\normalfont
    \begin{tabular}{@{}ccccccc@{}}
    \toprule
    \multirow{2}{*}{Model} & \multicolumn{2}{c}{Vulnerability Prediction} &  & \multicolumn{2}{c}{Clone Detection} &  \\ \cmidrule(l){2-7} 
     & Accuracy (\%) & Drop (\%) &  & Accuracy (\%) & Drop (\%) &  \\ \midrule
    CodeBERT (481 MB) & 61.82 & - &  & 96.20 & - &  \\
    \hdashline     BiLSTM$_{\text{\emph{soft}}}$ (7.5 MB) & 57.86 & 3.96  &  & 83.93  & 12.27  &  \\
    \toolname (3 MB) & \bf{59.44 (96.15\%)} &  \bf{2.38 (-39.90\%)} &  & \bf{95.43 (99.20\%)} & \bf{0.77 (-93.72\%)} &  \\ \midrule
    GraphCodeBERT (481 MB) & 61.38 & - &  & 96.62 & - &  \\
    \hdashline        BiLSTM$_{\text{\emph{soft}}}$ (7.5 MB) & 58.02 & 3.36 &  & 84.08 & 12.54 &  \\
    \toolname (3 MB) & \bf{59.99 (97.74\%)} & \bf{1.39 (-58.63\%)} &  & \bf{94.22 (97.52\%)} & \bf{2.4 (-80.86\%)}  &  \\ \bottomrule
    Average Maintained Accuracy/Improvements & \bf{96.95\%} & \bf{-49.27\%} &  & \bf{98.36\%} & \bf{-87.29\%}  &  \\ \bottomrule
    \end{tabular}
    \label{tab:rq1_1}
\end{table*}

\subsection{Evaluation Metric and Baseline}
\label{subsec:measurements}
A good model compression method is expected to (1) \emph{significantly reduce the model size} and (2) \emph{have minimal negative impacts on the model's performance} (e.g., accuracy). 
As suggested by Svyatkovskiy et al.~\cite{fastcompletion}, models of a 3 MB size are friendly to deploy in various devices, even some low-end hardware used to teach students to code. We compress the pre-trained models into 3 MB. Following the CodeXGLUE benchmark~\cite{lu2021codexglue} and Yang et al.~\cite{yang2022natural}, the accuracy is used as the evaluation metric on downstream tasks. We measure the prediction accuracy of tiny models obtained by \toolname and compute to what extent the accuracy decreases. 

Moreover, we consider a task-specific distillation method called \baseline introduced by Tang et al.~\cite{distillLSTM} as our baseline. \baseline is capable of compressing pre-trained models to 7.5 MB in our experiments, a size that is closed to the 3 MB expectation across the related literature.
We choose \baseline as our baseline for two reasons. On the one hand, the compressed models in their work also learn from the output logits of the pre-trained model; no other model outputs (e.g., the output of each encoder layer in pre-trained models) are required to perform knowledge distillation. Such settings are in line with ours. Second, \baseline is able to compress the pre-trained model to a very small size.

\baseline uses a shallow Bidirectional Long Short Term Memory (BiLSTM) network to distill a sizeable pre-trained model. More specifically, the student model has a fairly simple structure, consisting of a single-layer BiLSTM and two fully-connected layers. The training objective of \baseline is to minimize the mean squared error between the student model’s logits and the pre-trained model’s logits. Tang et al.~\cite{distillLSTM} also proposed using data augmentation techniques to harden the student model, which is to generate an amount of synthetic data by replacing words in a sentence based on the part-of-speech tags (i.e., grammatical properties of each word)~\cite{petrov2012universal}. Since source code does not have such tags, the data augmentation method is inherently unusable on compressed pre-trained code models. 


We use the code from their official repository\footnote{https://github.com/castorini/d-bert} in our experiments. Tang et al. provided information about all the hyperparameter settings of their method, except the vocabulary size. We follow all provided hyperparameter settings and keep the vocabulary size the same as the vocabulary size of our searched models.
Furthermore, to demonstrate the efficiency of compressed models, we focus on the inference latency (i.e., how long it takes the model to return a prediction for an input) each time the model is queried, which is important in a real-world deployment~\cite{fastcompletion, aye2020sequence}. 
We also take into account the time cost of the model compression process in \emph{Compressor}. The time cost consists of the time spent searching the parameters to simplify models and the time for knowledge distillation, demonstrating how long \toolname takes to compress the pre-trained models after fine-tuning.



\section{Experimental Results}
\label{sec:evaluation}

This section presents the evaluation results of our proposed approach and baselines. Our evaluation aims to answer the following research questions (RQ):
\begin{itemize}[leftmargin=*]
    \item \textbf{RQ1:} Can \toolname result in small accuracy loss when extremely compressing the pre-trained models?
    \item \textbf{RQ2:} How much efficiency improvement can the compressed models obtain?
    \item \textbf{RQ3:} How fast is \toolname in compressing pre-trained models?
\end{itemize}

\subsection{RQ1: Accuracy Loss of Compressor}

There is no free lunch in the world -- compressing the pre-trained models may penalize the original effectiveness of the models, i.e., having a lower prediction accuracy on downstream tasks. In this research question, we aim to show that \toolname can compress models to extremely small sizes with little loss of accuracy.

\vspace{0.15cm}
\noindent\textbf{Comparison with pre-trained models.} We use \toolname to compress the two pre-trained models of code into 3 MB each and investigate how much prediction accuracy the compressed models lose. Table~\ref{tab:rq1_1} presents the prediction accuracy of compressed models obtained by \toolname and the decrease in performance compared to the original pre-trained models in the $5^{th}$ and $8^{th}$ rows. 
We also report the results of \baseline in row 4 and row 7 to emphasize the improvements brought by \emph{Compressor}. 
We first analyze the accuracy loss of models compressed by our method compared to the accuracy of the original pre-trained models.
Our results show that the compressed models cause a negligible decrease in prediction accuracy when the model size is reduced to only 3 MB, which is only 0.6\% of the original volume. 
The compressed CodeBERT obtained by \toolname maintains 96.15\% of the original accuracy on the vulnerability prediction task. Meanwhile, \toolname achieves less accuracy degradation in compressing GraphCodeBERT, where the compressed model maintains 97.74\% of the original accuracy. 
On the clone detection task, compressed CodeBERT and GraphCodeBERT obtained by \toolname can achieve 99.20\% and 97.52\% of the original large models in terms of accuracy, respectively. These high fidelity accuracy results relative to the original model reveal that \toolname can significantly reduce model size with negligible accuracy loss, implying its effectiveness in compressing pre-trained models of code.

\vspace{0.15cm}
\noindent\textbf{Comparison with BiLSTM$_{\text{\emph{soft}}}$.}
We observe that \toolname can greatly outperform BiLSTM$_{\text{\emph{soft}}}$ on both two tasks, especially the clone detection task. 
Compared with the 7.5 MB BiLSTM$_{\text{\emph{soft}}}$, whose model size is more than twice that of the models obtained by \emph{Compressor}, the CodeBERT and GraphCodeBERT compressed by \toolname have reduced the accuracy loss by 93.72\% (i.e., (0.77-12.27)/12.27) and 80.86\%, respectively. 
On the vulnerability prediction task, the models obtained by \toolname also greatly reduce accuracy drops of BiLSTM$_{\text{\emph{soft}}}$: compressed CodeBERT reduces the accuracy drops of \baseline by 39.90\% and the compressed GraphCodeBERT reduces the accuracy drop by 58.63\%. These results demonstrate the significant superiority of \toolname over BiLSTM$_{\text{\emph{soft}}}$.

\vspace{0.15cm}
To sum up, when \toolname compresses the pre-trained models of code into 3 MB, which is only 0.6\% of the original model size, the degradation in accuracy is negligible on both two tasks. On average, the models compressed by \toolname can still maintain 96.95\% and 98.36\% of the accuracy of the original large models on both vulnerability prediction and clone detection tasks. Moreover, \toolname outperforms \baseline by 49.27\% and 87.29\% in reducing the accuracy loss of compressed models on both two tasks, respectively.

\begin{table}[t!]
    \centering
    \caption{The inference latency of original pre-trained and compressed models on different tasks. The percentage in parentheses is the relative improvements brought by \emph{Compressor}.}
    \renewcommand{\familydefault}{\sfdefault}\normalfont
    \begin{tabular}{@{}ccccc@{}}
    \toprule
    \multirow{2}{*}{Model} & \begin{tabular}[c]{@{}c@{}}Vulnerability\\ Prediction\end{tabular} &  & \begin{tabular}[c]{@{}c@{}}Clone\\ Detection\end{tabular} &  \\ \cmidrule(l){2-5} 
     & \multicolumn{3}{c}{Latency (ms)} &  \\ \midrule
    CodeBERT (481 MB) & 1507 &  & 2675 &  \\
    \toolname (3 MB) & \bf{347 (-76.97\%)} &  & \bf{625 (-76.64\%)} &  \\ \midrule
    GraphCodeBERT (481 MB) & 1209 &  & 1788 &  \\
    \toolname (3 MB) & \bf{429 (-64.52\%)} &  & \bf{326 (-81.77\%)} &  \\ \bottomrule
    Average Improvements & \bf{-70.75\%} &  & \bf{-79.21\%} & \\ \bottomrule
    \end{tabular}
    \label{tab:rq2}
    \end{table}

\subsection{RQ2: Efficiency of Compressed Models}

To upgrade the user experience, a plugin in IDEs or code editors is required to provide instantaneous assistance to developers so as to make the development process more efficient~\cite{aye2020sequence, fastcompletion}. In this research question, we investigate the efficiency of the compressed models obtained by \emph{Compressor}.
We analyze the inference latency when models make predictions, i.e., how much time a model takes to return a prediction for an input on average. The inference latency is related to two factors: the machine where the models run, and the input length (i.e., how many tokens are in the input). For fair comparisons, we use the same machine to run the models, which is mentioned in Section~\ref{sec:implementation}. We limit the models to only using 8 CPU cores to simulate running on a regular consumer-grade laptop. Usually, longer inputs take longer to compute. We set the input lengths to be the same as in experiments for RQ1, which is 400 for vulnerability prediction and 800 for clone detection, respectively. For each task, we randomly sample 100 examples from the test set. We query the models with the sampled 100 examples and calculate the average inference latency taken by the models. To reduce the effects of randomness, we repeat the experiments three times. The results are presented in Table~\ref{tab:rq2}. 

The inference latency of large pre-trained models is over a thousand milliseconds, while our compressed models have a latency of just a few hundred milliseconds.
On compressing CodeBERT, the average inference latency of the 3 MB models obtained by \toolname is 347 ms for vulnerability prediction and 625 ms for clone detection; the latter task takes longer inputs. 
Compared with the original pre-trained models, the inference latency of compressed models is significantly reduced by 76.97\% and 76.64\%, respectively. 
Setting sights to the compressed GraphCodeBERT, the compressed GraphCodeBERT is 64.52\% faster than the original pre-trained model on the vulnerability prediction task. On the clone detection task, the inference latency of the compressed GraphCodeBERT is 81.77\% less than the original pre-trained model. These results demonstrate the efficiency benefits that small models can potentially bring to real-world deployments, which is another key advantage of \emph{Compressor}.

\begin{table}[t!]
    \centering
    \caption{The time cost of fine-tuning and compressing pre-trained models on different tasks. The percentage in parentheses is the ratio of time spent by \toolname to fine-tuning time.}
    \renewcommand{\familydefault}{\sfdefault}\normalfont
    \begin{tabular}{@{}ccccc@{}}
    \toprule
    \multirow{2}{*}{Stage} & \begin{tabular}[c]{@{}c@{}}Vulnerability \\ Prediction\end{tabular} & & \begin{tabular}[c]{@{}c@{}}Clone \\ Detection\end{tabular} &  \\ \cmidrule(l){2-5} 
     & \multicolumn{3}{c}{Time Cost (min)} &  \\ \midrule
    CodeBERT Fine-tuning & 49  &  & 124 & \\ 
    \toolname & 13 (26.53\%)  &  & 47 (37.90\%) & \\ \midrule
    GraphCodeBERT Fine-tuning & 73  &  & 243 & \\  
    \toolname & 25 (34.25\%)  &  & 88 (36.21\%)  &\\ \bottomrule
    Average & \bf{30.39\%} &  & \bf{37.06\%} & \\ \bottomrule
\end{tabular}
\label{tab:rq3}
\end{table}

\subsection{RQ3: Time Cost of Compressor}
\label{subsec:rq3}
\toolname is a task-specific compression approach, which suggests that the pre-trained models to compress should be fine-tuned on a downstream task first. 
Therefore, applying \toolname during model development will inevitably introduce additional time costs. Since fine-tuning a large pre-trained model is already time-consuming, the subsequent compression process should not add too much overhead.
In this research question, we investigate the time spent by \toolname on compressing pre-trained models and aim to show that a well-performing tiny model can be obtained with much less time cost than fine-tuning large pre-trained models.

We repeat each experiment three times and count the time consumption of each stage, including fine-tuning and compressing the pre-trained models. We present the average time consumption in Table~\ref{tab:rq3}. Noted that we still consider 3 MB as the target size of compressed models.
Table~\ref{tab:rq3} displays the time consumption of fine-tuning and compressing the pre-trained models on different tasks. 
We observe that compressing the pre-trained models takes less time than fine-tuning the models on both tasks. The number inside parentheses is the ratio of compression time to the fine-tuning time. 
The time cost of model simplification and knowledge distillation is counted as a whole here. 
On the vulnerability prediction task, \toolname needs only 26.53\% and 34.25\% of fine-tuning time to compress CodeBERT and GraphCodeBERT, respectively. 
Although \toolname spends more time on the clone detection task, which needs to process a significantly larger dataset, as described in Section~\ref{sec:datasets}, it still only uses 37.90\% and 36.21\% of the fine-tuning time to compress CodeBERT and GraphCodeBERT, respectively. 
On average, \toolname only incurs 30.39\% and 37.06\% additional time to the fine-tuning time, which we believe to be reasonable. Note that this overhead is only incurred occasionally, e.g., the first time compressing the pre-trained models for deployment, and every time the model is on-demand to update for producing good results, which may be done monthly or yearly.

We also analyzed the efficiency of the GA-guided model simplification algorithm. 
Benefiting from the carefully-designed search space and fitness function, the GA-guided model simplification can return an appropriate model architecture fast that is then used as the student model in the subsequent knowledge distillation. To investigate the speed of the GA algorithm, we run the GA-guided search 10 times. On average, the GA-based algorithm spent 1.22 seconds searching a model architecture. The results further demonstrate the efficiency of \emph{Compressor}.

\section{Discussions}
\label{sec:discussion}
This section discusses whether the accuracy of compressed models can be boosted if we increase the model sizes. We point out the potential extensions of our proposed \emph{Compressor} and highlight the threats to validity and how to minimize these potential threats.

\subsection{The Impact of Compressed Model Sizes}
In our main experiment, we compress CodeBERT and GraphCodeBERT to a size of 3 MB, which is less than 1\% of the original model size. There may exist some other applications that have looser restrictions on the model size.
We discuss how the size of compressed models affects the model performance. 
We further experiment with two additional compression rates: 5\% and 10\%, to compress the models into around 25 MB and 50 MB. 

\begin{table}[t!]
    \centering
    \caption{The accuracy comparison of compressed models with different sizes. In the rows of Accuracy, the numbers in the parentheses correspond to the absolute improvement of accuracy compared with 3 MB models.}
    \renewcommand{\familydefault}{\sfdefault}\normalfont
    \begin{tabular}{@{}ccccc@{}}
    \toprule
    \multirow{2}{*}{Model} & \begin{tabular}[c]{@{}c@{}}Vulnerability\\ Prediction\end{tabular} &  & \begin{tabular}[c]{@{}c@{}}Clone\\ Detection\end{tabular} &  \\ \cmidrule(l){2-5} 
     & \multicolumn{3}{c}{Accuracy (\%)} &  \\ \midrule
    \toolname (3 MB) & 59.44 &  & 95.43 &  \\ \hdashline
    \toolname (25 MB) & 60.8 (+1.36) &  & 95.58 (+0.15) &  \\ 
    \toolname (50 MB) & \bf{61.31 (+1.87)} &  & \bf{96.30 (+0.87)} &  \\ \midrule
    \toolname (3 MB) & 59.99 &  & 94.22 &  \\ \hdashline
    \toolname (25 MB) & 60.32 (+0.33) &  & 95.65 (+1.43) &  \\
    \toolname (50 MB) & \bf{60.83 (+0.84)} &  & \bf{96.78 (+2.56)} &  \\ \bottomrule
    \end{tabular}
    \label{tab:dis}
    \end{table}

Table~\ref{tab:dis} displays the accuracy of pre-trained and compressed models of different sizes. On the vulnerability detection task, we observe that the performance of the 25 MB and 50 MB models (compressed from CodeBERT) increases by $1.36\%$ and $1.87\%$. Similarly, the two models compressed from GraphCodeBERT also have better performance: the accuracy increases by $0.33\%$ and $0.84\%$. 
On the clone detection task, the performance of the two models (compressed from CodeBERT) increases by $0.15\%$ and $0.87\%$. The two models compressed from GraphCodeBERT have $1.43\%$ and $2.56\%$ higher accuracy.
The results show that having a larger size can indeed improve compressed models' performance. It also further verifies the effectiveness of our proposed \emph{Compressor}. 

\subsection{Extensions of \emph{Compressor}}
\label{sec:extensions}

In Section~\ref{sec:evaluation}, we have demonstrated the advantages of \toolname in terms of high compression rates, low accuracy degradation, and efficiency. We now discuss the promising extensions that can be done to \emph{Compressor}.
We prototype \toolname as a model simplification tool that supports a limited number of popular fine-tuning/training pipelines that use HuggingFace Transformers. Its search space is constrained to the architecture of BERT-family models, due to their popularity and superior performance. However, the proposed method is promising to extend to support more pre-trained models with different architectures, e.g., CodeGPT~\cite{lu2021codexglue} and CodeT5~\cite{wang2021codet5}, with minimal effort to modify the way GFLOPs are calculated.




\subsection{Threats to Validity}

\subsubsection{Internal validity.}
In our experiments, we mainly focus on searching architecture-related hyperparameters to compress pre-trained models, but the accuracy of the compressed model can also vary under other remaining hyperparameters settings, e.g., input length, numbers of training epochs, etc.
To mitigate the threats, when fine-tuning pre-trained models, we keep the hyperparameters settings the same as described in~\cite{feng2020codebert,guo2021graphcodebert, lu2021codexglue}. We compare the performance of pre-trained models obtained in our study with results reported in the literature~\cite{lu2021codexglue} to show that our pre-trained models are properly trained. When compressing the model, we only modify the architecture-related hyperparameters as well as the learning rate, and the remaining hyperparameter settings are consistent with CodeBERT or GraphCodeBERT. In addition to these, another threat may lie in the implementations of \toolname and our scripts in the study. The authors have carefully checked the code before submission to reduce this threat.
We have also shared the code publicly for everyone to check and improve.

\subsubsection{External validity.}
The threats to external validity mainly lie in the used pre-trained models in our study.
In our experiments, we adopted two popular pre-trained models of code but they may not represent other pre-trained models such as C-BERT~\cite{CBERT}.
The current implementation of model simplification in \toolname only supports the models upholding BERT architecture. We plan to extend \toolname to compress other pre-trained models.

\subsubsection{Construct validity.}
The threats to construct validity mainly lie in the randomness of the measurements, especially when computing the time consumption. To reduce the impact of randomness in our study, we repeat each experiment multiple times and calculate averages. We use the commonly used metrics to evaluate the performance of our compression method, including the size of compressed models, inference latency, etc. Thus, we believe that this threat is minimal.
\section{Related Work}
\label{sec:rel_work}

This section describes the works that are related to this paper, including the pre-trained models of code, and existing compression techniques for pre-trained models.

\subsection{Pre-trained Models of Code}

Pre-trained models of code have recently become the state-of-the-art in the field of code intelligence and benefited a broad range of program understanding and generation tasks~\cite{feng2020codebert, guo2021graphcodebert, pmlr-v119-kanade20a, svyatkovskiy2020intellicode, wang2021codet5}. The pre-trained models can be categorized into three types: encoder-only, decoder-only, and encoder-decoder models.

Encoder-only models refer to pre-trained models that use only the encoder part of a Transformer model like BERT-family models. We have described two representatives of encoder-only models, CodeBERT~\cite{feng2020codebert} and GraphCodeBERT~\cite{guo2021graphcodebert}, in Section~\ref{subsec:pre-trained}. 
The two models have demonstrated good performance across multiple software engineering tasks, including API review~\cite{9825884}, Stack Overflow post analysis~\cite{9796213}, etc. 
There are some other encoder-only pre-trained models of code. Kanade et al.~\cite{CuBERT} employ the same model architecture and pre-training objective as BERT and pre-train a model called CuBERT on a large Python dataset. C-BERT~\cite{CBERT} also upholds BERT's powerful architecture and pre-training objective but is trained on the top-100 popular C language repositories in GitHub.
Both CuBERT and C-BERT can only process a single programming language, limiting their usage scenarios. No studies show that CuBERT and C-BERT can outperform CodeBERT and GraphCodeBERT, while the CodeXGLUE benchmark~\cite{lu2021codexglue} suggests that the latter two are among the state-of-the-arts, so this work investigates them in the experiment. 

Decoder-only models are pre-trained models that employ only the decoder part of a Transformer model. Decoder-only models are designed for generative tasks like code completion~\cite{9825794}. Svyatkovskiy et al.~\cite{svyatkovskiy2020intellicode} introduce GPT-C and MultiGPT-C, which achieve impressive performance in the code completion task. These two models are both variants of GPT-2~\cite{GPT-2}, but are trained on large corpora containing one or multiple programming languages. With the same model architecture and training objective, Lu et al.~\cite{lu2021codexglue} provide CodeGPT to help solve completion and generation problems. 

Encoder-decode models are pre-trained models that exploit the complete Transformer architecture in pre-training. Based on the BART~\cite{lewis-etal-2020-bart} architecture, PLBART~\cite{ahmad2021unified} is pre-trained on a large corpus of Java and Python functions with natural language descriptions collected from Github and StackOverflow. Wang et al.~\cite{wang2021codet5} adopt the T5~\cite{T5} model architecture to produce CodeT5 for both code understanding and generation tasks. CodeT5 proposes a novel identifier-aware pre-training objective that trains the model to distinguish which tokens are identifiers.

In our study, we focus on encoder-only pre-trained models since Clark et al.~\cite{Clark2020ELECTRA} provide an easy-to-use tool to calculate the FLOPs of these models. There has no such tool for other types of pre-trained models, so we leave them in the future investigation list.

\subsection{Compression of Pre-trained Models}

A prominent line of work is devoted to compressing large pre-trained models. As we discussed in Section~\ref{sec:intro}, there exist three groups of techniques: model pruning, model quantization, and knowledge distillation. 

Model pruning can be divided into unstructured pruning and structured pruning~\cite{10.1162/tacl_a_00413}. The former works by replacing some model parameters with zero. Gordon et al.~\cite{gordon-etal-2020-compressing} show that pruning 30\%-40\% of BERT's parameters does not affect the accuracy on downstream tasks. Structured pruning removes architectural components (e.g., network layers or attention heads) in the model. Fan et al.~\cite{Fan2020Reducing} propose to drop layers randomly during training, which allows for pruning some network layers at inference time without high accuracy degradation. Michel et al.~\cite{NEURIPS2019_2c601ad9} observe that a large percentage of attention heads in pre-trained models can be removed without significantly impacting performance. 
Unfortunately, model pruning cannot effectively compress pre-trained models of code to be 3 MB. Pre-trained models usually have an embedding table of around 150MB in addition to 12 (or 24) network layers. Even if all network layers in the pre-trained model are removed, the model still has a large embedding table. However, model pruning techniques can be used to further reduce the size of compressed models obtained by \emph{Compressor}, which is also a future work of ours.

Model quantization techniques convert the model's parameters from 32-bit floating-point numbers into low-bit fixed-point numbers. Zafrir et al.~\cite{zafrir2019q8bert} produce Q8BERT, which quantizes the network parameters and activations functions to 8-bit integers when fine-tuning BERT. Q8BERT can reduce the model size to 25\% of the original BERT size. Model quantization can reduce the model size. However, even if we convert all parameters into 1-bit binary values, the model can only be reduced to 32 times compared to the original size, which is still much larger than 3 MB.
Besides, a quantized model is not faster in inference speed or consumes less memory~\cite{10.1162/tacl_a_00413, zadeh2020gobo,pmlr-v139-kim21d}. 
Instead, specialized low-bit supporting hardware or optimized low-level processing libraries are required at the inference time. Therefore, we discard them from our experiments.

We have described the knowledge distillation techniques for pre-trained models in Section~\ref{sec:preliminaries} and introduced the \baseline as our baseline in Section~\ref{subsec:measurements}. We follow Svyatkovskiy et al.~\cite{fastcompletion} to set our targeted model size as 3 MB. Although several methods of compressing BERT can be adopted to compress CodeBERT or GraphCodeBERT, to our best knowledge, there are no studies that promise to compress a large pre-trained model of around 500 MB into such a tiny model of 3 MB. Most studies on compressing BERT by knowledge distillation can only result in models of sizes between 100 and 200 MB~\cite{distillbert, jiao2020tinybert, sun-etal-2019-patient,xu-etal-2020-bert}.
Chen et al.~\cite{10.5555/3491440.3491781} and Xu et al.~\cite{NASBERT} invariably adopt neural architecture search techniques to find a student model to compress BERT, and both obtain models of sizes between 20 and 40 MB. Zhao et al.~\cite{zhao-etal-2021-extremely} compress BERT to a 25 MB model with smaller vocabulary and hidden dimensions. Tang et al.~\cite{distillLSTM} adopt a shallow BiLSTM network to distill the large pre-trained models and obtain a 7.5 MB model with slight accuracy loss. Our study only compares with the work of Tang et al. as it is the only one that can compress pre-trained models close to 3 MB.
\section{Conclusion and Future Work}
\label{sec:conclusion}

  This paper proposes \emph{Compressor}, a novel approach that can compress the pre-trained models of code into tiny models sacrificing only negligible degradation in model accuracy. 
 \emph{Compressor} leverages a genetic algorithm (GA)-based strategy to maximize a model's Giga floating-point operations (GFLOPs) under a given model size constraint. 
 Then the small model learns from the larger pre-trained models using knowledge distillation.
  We evaluate \toolname with two state-of-the-art pre-trained models, i.e., CodeBERT and GraphCodeBERT, on the vulnerability prediction and clone detection tasks. 
  We use the proposed method to compress models to a size (3 MB), which is $160\times$ smaller than the original large models. The results show that compressed CodeBERT and GraphCodeBERT are $4.31\times$ and $4.15\times$ faster than the original model at inference time, respectively.
  More importantly, they maintain 96.15\% and 97.74\% of the original performance on the vulnerability prediction task. They even maintain higher ratios (99.20\% and 97.52\%) of the original performance on the clone detection task. 

  In the future, we plan to extend \emph{Compressor} to support models of more architectures and pre-trained models of code. We also plan to evaluate \emph{Compressor} using additional datasets and tasks beyond those considered in this paper.

\begin{acks}
This research/project is supported by the National Research Foundation, Singapore, under its Industry Alignment Fund – Pre-positioning (IAF-PP) Funding Initiative. Any opinions, findings and conclusions or recommendations expressed in this material are those of the author(s) and do not reflect the views of National Research Foundation, Singapore.
\end{acks}


\balance
\bibliographystyle{ACM-Reference-Format}
\bibliography{ref}

\end{document}